\documentclass[10pt,conference]{IEEEtran}
\usepackage{graphicx}
\usepackage{algorithm}
\usepackage{algorithmic}
\usepackage{amsmath}
\usepackage{amsfonts}
\usepackage{amssymb}
\newtheorem{theorem}{Theorem}

\newtheorem{definition}{Definition}

\begin{document}

\title{Delay and Throughput Optimal Scheduling for OFDM Broadcast Channels}

\author{Chan Zhou and Gerhard Wunder
\\ Fraunhofer German-Sino Lab for Mobile Communications (MCI), Heinrich-Hertz-Institut\\
  Einstein-Ufer 37, D-10587 Berlin (Germany)
  Email: \{zhou,wunder\}@hhi.fhg.de}

\maketitle \sloppy
%

\maketitle

\begin{abstract}
In this paper a scheduling policy is presented which minimizes the
average delay of the users. The scheduling scheme is investigated
both by analysis and simulations carried out in the context of
Orthogonal Frequency Division Multiplexing (OFDM) broadcast
channels (BC). First the delay optimality is obtained for a static
scenario providing solutions for specific subproblems, then the
analysis is carried over to the dynamic scheme. Furthermore
auxiliary tools are given for proving throughput optimality.
Finally simulations show the superior performance of the presented
scheme.
\end{abstract}

\section{Introduction}
The allocation of limited resources among users is a fundamental
problem in the design of next generation wireless systems. In
general, resource allocation problems can be formulated as some
kind of optimization problem where the objective is to
maximize/minimize some system performance measure under physical
layer as well as Quality of Service (QoS) constraints. One of the
most important performance measures of a communication systems is
the \emph{total system throughput} and therefore it is often
considered as the objective of the optimization problem. In a
queueing system with random packet arrival, the throughput can be
considered as the maximal possible offered load without violating
the stability of queues. A scheduling policy is called
\emph{throughput-optimal}, if it can keep the queues stable
whenever any other feasible scheduling scheme can stabilize the
queues. It was shown that there exist several queue-length-based
scheduling schemes which achieve throughput optimality
\cite{Tassiulas:1992,Andrews:2001,Shakkottai:2001,Seong:2006}.

However, since the stability of a queueing system only guarantees
that the queue lengths do not grow without bounds, but by no means
indicate how long the queue length will be, the next step in the
performance optimization is to keep the queue lengths as short as
possible so that the queueing delay is minimized. It was shown in
\cite{YehCoh:2004,ISITA06} that the
Longest-Queue-Highest-Possible-Rate (LQHPR) policy which maximizes
the queue-weighted sum of rates is throughput-optimal and is
strongly delay-optimal for the multiple-access channel. The
necessary condition for its delay optimality is the symmetry both
in the fading channels and in the packet arrival rates. However,
for the BC the LQHPR is not delay-optimal even with symmetry
assumptions. Seong et al. introduced in \cite{Seong:2006} another
throughput-optimal scheduling called Queue Proportional Scheduling
(QPS) which provides superior delay and fairness properties for
the BC compared to LQHPR.

Generally, the aforementioned policies are based on the same class
of optimization problems: Maximizing the sum of rates weighted
with different parameters, i.e. queue length, delay, etc. The
solution of the optimization problem always corresponds to some
boundary point of the channel capacity region. Since the capacity
region of OFDM BC is completely achieved with Costa Precoding and
the optimal power allocation and precoding order for the weighted
sum rate maximization problem can be efficiently solved
\cite{EUSIPCO06}, all results can also be easily extended to OFDM
BC systems.

In this paper, we analyze characteristics of all
throughput-optimal scheduling policies and show that they can be
formulated as weighted sum rate maximization problems differing
only in the choice of the weight factors. Further, the weight
factors are independent of the current channel states, hence
cross-layer optimization problems, which usually involve the
optimization over system parameters in medium access control (MAC)
layer and physical layer, can be clearly separated into two steps:
1. Finding the optimal weight factors according to the MAC layer
parameters. 2. Solving the weighted sum rate maximization problem.
Then we introduce an iterative algorithm to calculate the weight
factors that are optimal with respect to the average delay in OFDM
BC channels. Here the average delay is defined as the average
waiting time of each bit in the queues.

The rest of this paper is organized as follows. Section II
presents the system model. The throughput optimality is discussed
in Section III. In Section IV we introduce our delay-optimal
scheduling policy and evaluate results in Section V. Finally, we
conclude in Section VI.

\section{System Model}
\subsection{Physical layer}
We assume an OFDM BC with $M$ users, $K$ subcarriers, and a short
term sum power constraint $\bar{P}$
\[
{\sum_{m,k=1}^{M,K}\mathbb{E}\{|x_{m,k}|^{2}\}=\sum_{m,k=1}^{M,K}p_{m,k}\leq\bar{P}},
\]
where $x_{m,k}$ is the signal transmitted to user $m\in\mathcal{M}%
=\{1,...,M\}$ on subcarrier $k\in\mathcal{K}=\{1,...,K\}$ with
power ${p_{m,k}}$ and $\mathbb{E}\{.\}$ stands for the expectation
operator. Then, the system equation for each user on each
subcarrier can be
written as%
\begin{equation}
   y_{m,k}=h_{m,k}\sum_{j\in\mathcal{M}}\,x_{j,k}+n_{m,k},\quad m\in
   \mathcal{M},k\in\mathcal{K},
\end{equation}
where $y_{m,k}$ is the signal received by user $m$ on subcarrier
$k$, $n_{m,k}\sim\mathcal{CN}(0,\sigma^{2})$ is circular symmetric
additive white Gaussian noise with variance $\sigma^{2}$. Let
$\mathbf{h}=[h_{1,1}%
,\ldots,h_{1,K},h_{2,1},\ldots,...h_{M,K}]^{T}$ denote the stacked
vector of channel coefficients.  We assume that these channel
coefficients are related to a standard time-varying multipath
model where the channel is approximately constant over the OFDM
symbol. Furthermore, we assume that Costa Precoding is performed
at the base station having full non-causal knowledge of all
messages to be transmitted. Let $\pi\in\Pi$ be an arbitrary
encoding order from the set of all $M!$ possible encoding orders,
such that user $\pi(1)$ is encoded first, followed by user
$\pi(2)$ and so on. Then the rate of user $\pi(m)$ can be
expressed as
\begin{equation}
   \tilde{r}_{\pi(m)}=\sum_{k=1}^{K}\log\left(  1+\frac{g_{\pi(m),k}
       p_{\pi(m),k}}{1+g_{\pi(m),k}\sum_{n<m}p_{\pi(n),k}}\right)
   \label{rate_bc}%
\end{equation}
with $g_{m,k}=|h_{m,k}|^2/\sigma^2$ being the channel gain of user
$m$ on subcarrier $k$ and $p_{m,k}$ being the allocated power.
The instantaneous capacity region of the OFDM BC under a given sum
power constraint $\bar{P}$ is given by
\begin{equation}
\label{eq:def_C_inst}
   \mathcal{C}(\mathbf{h},\bar{P} )\equiv\bigcup\limits_{\substack{\pi\in\Pi
       \\\sum_{m,k=1}^{M,K}p_{m,k}\leq\bar{P}}}\left\{  \mathbf{r}:r_{\pi(m)}%
     \leq\tilde{r}_{\pi(m)}\;,m\in\mathcal{M}\right\}
\end{equation}
where $\tilde{r}_{\pi(m)}$ is defined in equation (\ref{rate_bc})
and $\mathbf{r}$ denotes the vector of rates. Now, the
\emph{ergodic
  capacity region} $\mathcal{C}_{erg}(\bar{P})$ is defined as the set of
achievable rates averaged over the channel realizations subjected
to the short term sum power constraint $\bar{P}$:
\begin{equation}
\label{eq:def_C_erg}
   \mathcal{C}_{erg}(\bar{P})\equiv\bigcup\limits_{\substack{\pi\in\Pi
       \\\sum_{m,k=1}^{M,K}p_{m,k}\leq\bar{P}}}\left\{  \mathbf{r}:r_{\pi(m)}%
     \leq \mathbb{E}_{\mathbf{h}} \left \{ \tilde{r}_{\pi(m)} \right \},m\in\mathcal{M}\right\}
\end{equation}

\subsection{Medium access control layer}

Assuming that the transmission is time-slotted, data packets
arrive randomly at the MAC and a buffer with finite length is
reserved to store the incoming data for each user
$m\in\mathcal{M}$. Simultaneously the data is read out from the
buffers according to the system state, i.e., the random fading
realization and the current queue lengths. Thus, the system can be
modeled as a queueing system with random processes reflecting the
arrival and the departure of data packets.

Denoting the buffer state of the $m$-th buffer in time slot
$n\in\mathbb{N}$ by $q_{m}\left(  n\right)  $ and arranging all
buffer states in the vector $\mathbf{q}(n)\in\mathbb{R}_{+}^{M}$
the evolution of the queue system can be written as
\begin{equation}
   \mathbf{q}\left(  n+1\right)  =\left[  \mathbf{q}\left(  n\right)
     -\mathbf{r}\left(  n\right)  \right]  ^{+} +\mathbf{a}\left(
n\right),
\end{equation}
where $[{x}]_m^{+} = \max\{0,x_{m}\}$, $\forall m \in \mathcal{M}$
 and $\mathbf{a}\left(n\right) \in\mathbb{R}_{+}^{M}$ is a random
vector denoting the data arrival process. The random vector
$\mathbf{r}\left( n\right) \in\mathbb{R}_{+}^{M}$ describes the
rates asserted to the individual users according to a specific
scheduling policy. Supposing that a scheduling policy is a mapping
\[
\mathcal{P}: \mathbb{C}^{M \times K} \times \mathbb{S} \rightarrow
\mathbb{R}^{M}_{+},
\]
which decides the rate allocation depending on the current channel
fading state $\mathbf{h} \in \mathbb{C}^{M \times K}$ and the MAC
layer system state $\mathbf{w} \in \mathbb{S}$. $\mathbf{w}$
summarizes the current and past information that is acquirable at
the base station and relevant for the optimization, e.g., the
current queue length $\mathbf{q}$, the average previous arrival
rate $\bar{\mathbf{a}}$ and average previous transmit rate
$\bar{\mathbf{r}}$, etc. The rate allocation according to the
scheduling policy $\mathcal{P}$ is denoted as
$\mathbf{r}^{\mathcal{P}}(\mathbf{h},\mathbf{w})$.

Note that the process is reminiscent of random walk on the half
line (with dependent increments) where we have rigorously used an
uncountable state space formulation. Since the random variables
$\mathbf{a}(n)$ are sampled at a given time interval $T$ from $M$
independent random processes they are independent. Denoting the
mean of the packet arrival rate of user $m$ as $\lambda_{m}$ and
the constant packet size as $s_{m}$, the expected bit arrival rate
for user $m$ is given by $\rho_{m}=s_{m}\lambda_{m}$. On the other
hand the random vector $\mathbf{r}\left( n\right) $ depends on the
buffer and channel state.

\section{System Stability and Throughput-optimal Scheduling Policies}

\subsection{Definition of system stability and throughput optimality}

First we investigate the maximum possible offered system load
without violating stability. There exist several definitions of
stability. In this paper we use the definition of the \emph{strong
stability}, which implies also \emph{weak stability} and
\emph{nonevanescence} of the queueing system.

\begin{definition}
The queueing system is strongly stable, if
\begin{equation}
\limsup_{n \rightarrow + \infty} \mathbb{E} \left\{ q_i(n) \right
\} < +\infty, \forall i \in \mathcal{M}.
\end{equation}
\end{definition}

In the sense of the stability definition, we call the set of
expected arrival rates $\mathbf{\rho}$ \emph{stabilizable} by a
specific scheduler the \emph{throughput region} of the scheduling
policy $\mathcal{P}$. A scheduling policy is
\emph{throughput-optimal} if it stabilizes the system whenever any
other scheduling policies can stabilize the system. If the arrival
rate $\boldsymbol{\rho} \notin C_{erg}(\bar{P})$ and the fading
gains can be practically upperbounded by some constants, then it
is impossible to stabilize the system, even if the policy is
non-stationary and it has knowledge of the future events
\cite{EUSIPCO06}. Therefore, we can define a
\emph{throughput-optimal} scheduling policy as a policy, which
keeps the system stable for any arrival rate whose expected value
$\boldsymbol{\rho}$ lies in the ergodic capacity region. For
example, using Lyapunov drift technique the LQHPR scheduler can be
proven to be throughput optimal.

\subsection{Characterization of throughput-optimal scheduling
policies}

For a given channel state, the rate vector allocated with a
throughput-optimal policy is always a boundary point of the
instantaneous capacity region $\mathcal{C}(\mathbf{h},\bar{P})$.
Therefore, any throughput-optimal scheduling policy can be
formulated as the optimization problem

\begin{equation}
\label{eq:weighted_sum_rate}
\mathbf{r}^{\mathcal{P}}(\mathbf{h},\mathbf{w}) =
\underset{\mathbf{r} \in
\mathcal{C}\left(\mathbf{h},\bar{P}\right)}{\arg\max}
\boldsymbol{\mu}^T \cdot \mathbf{r},
\end{equation}
where $\boldsymbol{\mu}$ is the normal vector of the boundary
surface at the allocated rate point. Since the boundary surface of
the OFDM BC channel is continuous and differentiable, the rate
allocation $\mathbf{r}^{\mathcal{P}}(\mathbf{h},\mathbf{w})$ can
be uniquely characterized with the normal vector
$\boldsymbol{\mu}$ on the capacity region.

Furthermore, for any given $\boldsymbol{\mu}$ the power and rate
allocation problem (\ref{eq:weighted_sum_rate}) can be efficiently
solved \cite{EUSIPCO06}. Hence, we can use the vector
$\boldsymbol{\mu}$, which is also called weight vector in the
optimization problem (\ref{eq:weighted_sum_rate}), to characterize
the scheduling decision, instead of using power and rate
allocation directly.

In the following we show some properties of the weight vector
$\boldsymbol{\mu}$.

\begin{theorem}
The weight vector $\boldsymbol{\mu}$ which characterizes a
throughput-optimal scheduling policy is independent of the current
fading state $\mathbf{h}$. \label{theorem:independence}
\end{theorem}

\begin{proof}
We choose arbitrarily a weight vector $\boldsymbol{\mu}^*$
corresponding to a fixed boundary point of the ergodic capacity
region, hence $\boldsymbol{\mu}^*$ is independent of the
instantaneous channel state. Then we denote
$\boldsymbol{\mu}_{\mathbf{h}}$ the weight vector determined by a
scheduling policy $\mathcal{P}$. We have
\begin{align}
    & \boldsymbol{\mu}^{*T} \cdot \mathbb{E}_{\mathbf{h}}\left\{ \mathbf{r}^\mathcal{P}\left(\mathbf{h},\mathbf{w} \right ) \right\} \\
    = & \boldsymbol{\mu}^{*T} \cdot \mathbb{E}_{\mathbf{h}}\left\{ \arg\max _{\mathbf{r} \in  \mathcal{C}\left(\mathbf{h},\bar{P}\right)} \boldsymbol{\mu}_{\mathbf{h}}^T \cdot
    \mathbf{r} \right\} \\
    = & \mathbb{E}_{\mathbf{h}}\left\{ \boldsymbol{\mu}^{*T} \cdot \arg\max _{\mathbf{r} \in  \mathcal{C}\left(\mathbf{h},\bar{P}\right)} \boldsymbol{\mu}_{\mathbf{h}}^T \cdot
    \mathbf{r} \right\} \\
\label{ineq:max_u_r}
    \leq & \mathbb{E}_{\mathbf{h}}\left\{ \max _{\mathbf{r} \in  \mathcal{C}\left(\mathbf{h},\bar{P}\right)} \boldsymbol{\mu}^{*T} \cdot
    \mathbf{r} \right\},
\end{align}
The equality holds only if $\boldsymbol{\mu}_{\mathbf{h}} =
\boldsymbol{\mu}^*$ and the boundary point is achieved by the
corresponding scheduler, otherwise the scheduling policy gives a
rate vector in the interior of the ergodic capacity region. If the
expected rates of the arrival process equals the rates on the
boundary point, there must be some user $i$ who has
$\mathbb{E}_{\mathbf{h}}\left\{
r_i^\mathcal{P}\left(\mathbf{h},\mathbf{w} \right) \right \} <
\rho_i$ and its queue expands infinitely.
\end{proof}

Following the result in Theorem \ref{theorem:independence}, we can
define the weight vector
$\boldsymbol{\mu}^{\mathcal{P}}(\mathbf{w})$ of a
throughput-optimal policy as a function only determined by the MAC
layer system state $\mathbf{w}$. In this way, the classical
\emph{cross-layer} optimization problem can be separated into two
parts: Finding the optimal weight vector $\boldsymbol{\mu}$
according to the MAC layer parameters; solving the rate and power
allocation problem (\ref{eq:weighted_sum_rate}) on the physical
layer with the giving weight vector. Since the second part can be
efficiently solved on the physical layer, the scheduling design
problem reduced to find the optimal weight vector for the
optimization problem.

\section{Delay-optimal scheduling policy}
So far we characterize the class of throughput-optimal scheduling
policies. Since the stability definition doesn't restrict the
explicit length of the queues, even throughput-optimal policies
have different delay performance. In the following we study the
scheduling policy minimizing the average bit delay $\bar{D}$,
which is defined as:
\begin{equation}
\bar{D} = \frac{1}{M} \sum^{M}_{i=1} D_i = \frac{1}{M}
\sum^{N}_{n=1} \sum^{M}_{i=1} \frac{q_i(n)}{\bar{a}_i}.
\end{equation}
It can be regarded as the extension of the common definition of
the queueing delay in \cite{Kleinrock_Queueing_Systems_I}. $N$ is
the length of the observation time window and $\bar{a}_i$ is the
average bit arrival rate for the user $i$ in the time window.

\subsection{Delay-optimal scheduling policy for a static channel}
\label{sec:Delay_optimal_static} We consider first the delay
optimization problem for a static channel $\mathbf{h}$ and the
initial buffer states $\mathbf{q}(n=1)$. We denote the previous
average arrival rates as $\bar{\mathbf{a}}$ and assume there is
\emph{no} packet arriving after $n=0$. Further we choose the
length of observation time window $N$ with $q_i(N) = 0, \; \forall
i \in \mathcal{M}$ so that the buffers are completely emptied
within the time window.

Thus the delay-optimal scheduling policy can be written as the
solution of the optimization problem
\begin{align}
\label{problem:min_delay_static}
& \min \sum^{M}_{i=1} D_i \nonumber \equiv \min \sum^{N}_{n=1} \sum^{M}_{i=1} \frac{q^n_i}{\bar{a}_i}\\
& s.t. \qquad q^{n+1}_i = q^n_i - r^n_i \\
\nonumber & \quad \qquad \mathbf{r^n} \in \mathcal{C}\left(\mathbf{h},\bar{P}\right)\\
\nonumber & \quad \qquad q^n_i - r^n_i \geq 0, \qquad \forall i
\in \mathcal{M}, n \in [1,...,N],
\end{align}
where $q^n_i$, $r^n_i$ denote the queue length and transmit rate
of user $i$ in time slot $n$. For convenience we also use the
superscript to denote the time slot in the following. Extending
the problem (\ref{problem:min_delay_static}) in each queue state
$\mathbf{q}^n$ we have the equivalent optimization problem

\begin{align}
\label{problem:min_delay_static2} & \min \sum^{N}_{n=1} \left(
\sum^{M}_{i=1} \frac {q^1_i}{\bar{a}_i} - \sum^{M}_{i=1} \left(
N-n
\right) \frac{r^n_i}{\bar{a}_i} \right) \nonumber \\
& s.t. \qquad \mathbf{r^n} \in \mathcal{C}\left(\mathbf{h},\bar{P}\right)\\
\nonumber & \quad \qquad q^1_i - \sum^{n}_{t=1}r^t_i \geq 0,
\qquad \forall i \in \mathcal{M}, n \in [1,...,N]
\end{align}

The Lagrangian function is
\begin{align}
& L(\mathbf{r}^n,\boldsymbol{\lambda}^n) \\
\nonumber  = & \sum^N_{n=1} \sum^{M}_{i=1} \frac{q^1_i}{\bar{a}_i}
- \sum^N_{n=1} \sum^{M}_{i=1} (N-n) \frac{r^n_i}{\bar{a}_i} -
\sum^N_{n=1} \sum^{M}_{i=1} \lambda^n_i \left( q^1_i - \sum^n_t
r^t_i\right)
\end{align}

Denote $\eta_i^* = N-\bar{a}_i\sum_{t=1}^{N}\lambda_i^t$, we get
the optimal $\mu_i^n$ with
\begin{equation}
\label{eq:def_mu} \mu_i^n = \left \{ \begin{tabular}{ll}
 $\frac{\eta_i^* - n + 1}{\bar{a}_i} \;$ &$ n \leq \eta^*$ \\
 $0$ \; &$n > \eta^*$
\end{tabular}
\right.
\end{equation}
and the delay-optimization problem is transformed into
\begin{align}
\label{problem:min_delay_mu} & \max \sum^{N}_{n=1} \sum^{M}_{i=1}
\boldsymbol{\mu}^{nT} \cdot \mathbf{r} \nonumber \\
 s.t. & \qquad \mathbf{r^n} \in \mathcal{C}\left(\mathbf{h},\bar{P}\right)
\end{align}
which can be solved easily.

The $\eta_i^*$ in (\ref{eq:def_mu}) can be obtained with a
iterative approach given in Algorithm \ref{algorithm:1}.

\begin{algorithm}
\caption{Idle State Prediction Algorithm}\label{algorithm:1}
\begin{algorithmic}
\STATE{\bf(1)} Set $\mu_i^{(0)} = \frac{1}{\bar{a}_i}$ and
calculate $\mathbf{r}^{(0)} = \arg\max_{\mathbf{r} \in
\mathcal{C}_{\mathbf{h}}} \boldsymbol{\mu}^{(0)T} \cdot
\mathbf{r}$.

\STATE{\bf(2)} Initialize the length of non-idle state
$\eta^{(0)}_i = \min_{i \in \mathcal{M}} \frac{q^1_i}{r^{(0)}_i}$.

\STATE{\bf(3)} Set the order $\pi$ so that
$\frac{q^1_{\pi(1)}}{r^{(0)}_{\pi(1)}} \geq
\frac{q^1_{\pi(2)}}{r^{(0)}_{\pi(2)}} \geq ... \geq
\frac{q^1_{\pi(M)}}{r^{(0)}_{\pi(M)}}$.

\STATE{\bf(4)} Set $t = 0$.

\REPEAT

\STATE{\bf(5.1)} Set $\boldsymbol{\eta}^{(t+1)} =
\boldsymbol{\eta}^{(t)}$

\FOR{$i=1$ to $M$}

\STATE{\bf(5.2.1)} $\boldsymbol{\eta}^* =
\boldsymbol{\eta}^{(t+1)}$

\REPEAT

\STATE{\bf(5.2.2.1)} Increase $\eta_{\pi(i)}^*$. Solve the
maximization problem (\ref{problem:min_delay_mu}) and calculate
the evolution of the queue state.

\IF{$q_{\pi(i)}^{\lceil \eta^*_{\pi(i)} \rceil} \geq 0$}

\STATE $ \eta_{\pi(i)}^{(t+1)} = \eta_{\pi(i)}^*$

\ENDIF

\UNTIL {$q_{\pi(i)}^{\lceil \eta^*_{\pi(i)} \rceil} < 0$}

\ENDFOR

\STATE $t = t + 1$

\UNTIL {$\eta_i^{(t)}-\eta_i^{(t-1)} < \epsilon, \quad \forall i
\in \mathcal{M}$}

\STATE{\bf(3)} $\boldsymbol{\eta}^* = \boldsymbol{\eta}^{(t)}$
\end{algorithmic}
\footnotesize $\lceil \eta \rceil$ denotes the smallest integer
larger than $\eta$. \\
$\epsilon$ is the predefined error tolerance of $\eta$.
\normalsize
\end{algorithm}

\begin{theorem}
$\boldsymbol{\eta}^{(t)}$ obtained in Algorithm \ref{algorithm:1}
converges to the to $\boldsymbol{\eta}^*$ which gives the optimal
$\mu_i^n$ for the delay-optimization problem
(\ref{problem:min_delay_static}).
\end{theorem}

\begin{proof}
In any time slot $n > \eta_i^*$, we have $\mu_i^n = 0$ which means
the buffer of $i$-th user is empty at the $n$-th time slot. In any
$n \leq \eta_i^*$, the $i$-th buffer must be non-empty. Therefore
if $\boldsymbol{\eta}^{(t)}=\boldsymbol{\eta}^*$, we have
$q_i(\lceil \eta^*_i \rceil) = 0$, $\forall i \in \mathcal{M}$ and
the algorithm stops at the optimum.

For two users $i,j \in \mathcal{M}$, if $\eta^*_i > \eta^*_j$, the
optimal weight factors
\[
\frac{\mu^n_i}{\mu^n_j} = \frac{\frac{\eta_i^*-n+1}{\bar{a}_i}}{
\frac{\eta_j^*-n+1}{\bar{a}_j}} > \frac{\frac{1}{\bar{a}_i}}{
\frac{1}{\bar{a}_j}} = \frac{\mu^{(0)}_i}{\mu^{(0)}_j} \qquad n
\in [1,...,\lceil \eta ^*_i \rceil]
\]
and
\begin{equation}
\label{ineq:r_opt_r_0} \frac{r^{n}_i}{r^{n}_j} >
\frac{r^{(0)}_i}{r^{(0)}_j} \qquad n \in [1,...,\lceil \eta^*_i
\rceil]
\end{equation}
follows.

From (\ref{ineq:r_opt_r_0}) we have
\[
\frac{r^{(0)}_i}{r^{(0)}_j} <
\frac{\sum_{n=1}^{\lceil\eta^*_i\rceil}r^{n}_i}{\sum_{n=1}^{\lceil\eta^*_i\rceil}r^{n}_j}
= \frac{q^1_i}{q^1_j}
\]

Therefore, $\eta^*_{\pi(i)} \geq \eta^*_{\pi(j)}$ holds if $i <
j$. Further, since $\eta_i^{(0)} =
\frac{q^1_{\pi(M)}}{r^{(0)}_{\pi(M)}} \leq \eta^*_{\pi(i)}$,
$\forall i \in \mathcal{M}$, the initial state $\eta_i^{(0)} \leq
\eta_i^{*}$, $\forall i \in \mathcal{M}$. In each iteration step,
if the optimum $\boldsymbol{\eta}^*$ is not achieved,
$\eta^{(t+1)}_i$ can always be increased so that $\eta_i^{(t+1)}
> \eta_i^{(t)}$, $\forall i \in \mathcal{M}$. Hence, the convergence of the algorithm is
proven.
\end{proof}

\subsection{Delay-optimal scheduling for dynamic channels}
It is worth noting that if channel state $\mathbf{h}$ varies over
time and the base station has the knowledge of each channel state
in advance, the algorithm in previous subsection can also be used
in this case with some modification. However, in reality the base
station has usually only the current channel state information and
the statistical knowledge of the channel. Further the packet
arrival process is non-ergodic and cannot be predicted. In order
to avoid the possible infinite queueing delay, the delay-optimal
policy must also be throughput-optimal, so that the queue state is
kept stable for any expected arrival rate $\boldsymbol{\rho}$
inside the ergodic capacity region.

If no new packet arrives after the time slot $n=0$, the expected
delay for a given policy $\mathcal{P}$ is
\begin{align}
\label{problem:min_delay_dynamic} & \mathbb{E} \left \{
\sum^{N}_{n=1} \sum^{M}_{i=1} D^n_i  \right \}
\nonumber \\
= & \mathbb{E} \left \{ \sum^{N}_{n=1} \left( \sum^{M}_{i=1} \frac
{q^1_i}{\bar{a}_i} - \sum^{M}_{i=1} \left( N-n \right)
\frac{r^{\mathcal{P}n}_i}{\bar{a}_i} \right) \right \},
\end{align}
where $r^{\mathcal{P}n}_i$ is the rate allocated by the policy
$\mathcal{P}$ for the $i$-th user at $n$-th time slot. From
Theorem \ref{theorem:independence} we know that if $\mathcal{P}$
is a throughput-optimal policy, then
\begin{equation}
\mathbb{E}\left\{ \mathbf{r}^{\mathcal{P}} \right \} =
\arg\max_{\mathbf{r} \in \mathcal{C}_{erg}(\bar{P})}
(\boldsymbol{\mu}^{\mathcal{P}})^T \cdot \mathbf{r},
\end{equation}
 where
$\boldsymbol{\mu}^{\mathcal{P}}$ is independent of the current
channel state. Hence the optimization problem is equivalent to
\begin{align}
\label{problem:min_delay_dynamic2} & \min \sum^{N}_{n=1} \left(
\sum^{M}_{i=1} \frac {q^1_i}{\bar{a}_i} - \sum^{M}_{i=1} \left(
N-n
\right) \frac{\tilde{r}^n_i}{\bar{a}_i} \right) \nonumber \\
& s.t. \qquad \mathbf{\tilde{r}^n} \in \mathcal{C}_{erg}\left(\bar{P}\right)\\
\nonumber & \quad \qquad q^1_i - \sum^{n}_{t=1}r^t_i \geq 0,
\qquad \forall i \in \mathcal{M}, n \in [1,...,N].
\end{align}

Then the optimization problem can be solved using algorithm
\ref{algorithm:2}.

\begin{algorithm}
\caption{Delay Optimal Scheduling}\label{algorithm:2}
\begin{algorithmic}
\FOR {each time slot $n$}

\STATE{\bf{1}} Calculate the pervious average arrival rate
$\bar{\mathbf{a}}$

\STATE{\bf{2}} Calculate $\boldsymbol{\eta}^*$ according to
$\bar{\mathbf{a}}$ and current queue state $\mathbf{q}$ using
algorithm \ref{algorithm:1}, where the static channel region
$\mathcal{C}(\mathbf{h},\bar{P})$ is replaced with the ergodic
capacity region $\mathcal{C}_{erg}(\bar{P})$.

\STATE{\bf{3}} Calculate the current weight vector
$\boldsymbol{\tilde{\mu}}^1$ according to equation
(\ref{eq:def_mu}).

\STATE{\bf{4}} Calculate the current rate allocation
\begin{equation} \mathbf{r}^* = \arg\max_{\mathbf{r} \in
\mathcal{C}(\mathbf{h}^n, \bar{p})} (\boldsymbol{\tilde{\mu}}^1)^T
\cdot \mathbf{r},
\end{equation}
where $\mathbf{h}^n$ is the current channel state.

\ENDFOR
\end{algorithmic}
\end{algorithm}

In the system with new packet arrival, the weight vector
$\boldsymbol{\tilde{\mu}}^1$ should be recalculated according to
the new queue state and the rate allocation is determined with
$\boldsymbol{\tilde{\mu}}^1$ and current channel state
$\mathbf{h}$.

\begin{theorem}
The proposed scheduling policy keeps the queue lengths finite for
all arrival processes with expected arrival rates inside the
capacity region.
\end{theorem}

Sketch of the proof: It is easy to show that $\tilde{\mu}_i^1$
monotonically increases from $0$ to infinity if $q_i$ grows from
$0$ to infinity, for all $i \in \mathcal{M}$. Therefore, the
expected transmit rate vector $\mathbb{E}\{\mathbf{r}^*\}$
converges to a boundary point $\hat{\mathbf{r}}$ where
$\frac{q_i}{q_j} = \frac{\rho_i - \hat{r}_i}{\rho_j - \hat{r}_j}$,
$\forall i,j \in \mathcal{M}$. Since $\boldsymbol{\rho}$ lies
inside the ergodic region, we have $\hat{r}_i \geq \rho_i$,
$\forall i \in \mathcal{M}$ and the theorem follows.

\section{Numerical Evaluations}
\label{section:evaluations}

\begin{figure}[h]
\begin{center}
\includegraphics[width=\linewidth]{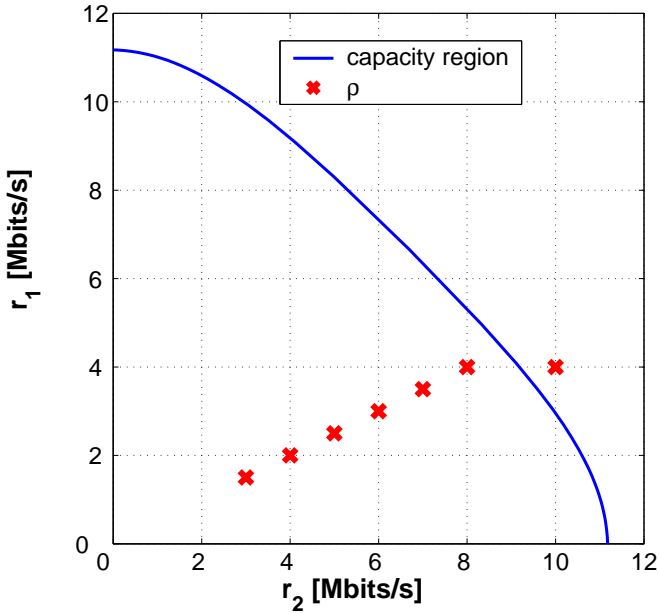}
\end{center}
\caption{{\footnotesize {Ergodic capacity region for 2 users}}}%
\label{fig_caps}%
\end{figure}

\begin{figure}[h]
\begin{center}
\includegraphics[width=\linewidth]{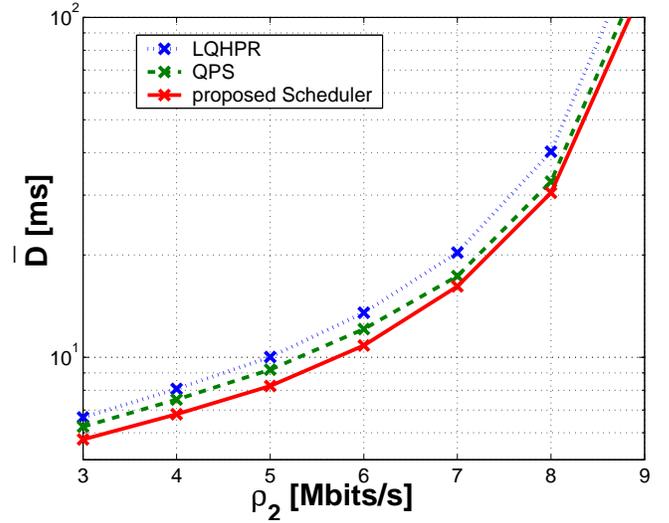}
\end{center}
\caption{{\footnotesize {Average bit delay}}}%
\label{fig_bound}%
\end{figure}

We compare our scheduler with the LQHPR scheduler in
\cite{Tassiulas:1992} and the QPS in \cite{Seong:2006} in a
two-user scenario. The OFDM system has 250 subcarriers and an
entire bandwidth of 2.5MHz. The multipath channel has i.i.d block
fading model and the length of fading block $T$ was assumed to be
0.1ms. For an average transmit SNR of 15dB the ergodic capacity
region was shown in Fig.\ref{fig_caps}. Having chosen $\rho _{1} =
[1.5, 2, 2.5, 3, 3.5, 4, 4]$ Mbits/s and $\rho_{2} = [3, 4, 5, 6,
7, 8, 10]$ Mbits/s the average bit delay is shown in
Fig.\ref{fig_bound}. It can be seen that for the arrival rate
outside the capacity region, the queueing delay becomes extremely
long. For the arrival rate inside the capacity region, the
introduced scheduling policy has superior delay performance
compared to the other two scheduling policies.

\section{Conclusion}
We have provided a throughput and delay optimal scheduling policy
for OFDM BC channels. Simulation results show that the average bit
delay is significantly reduced with the introduced scheduling
policies.

Since the weighted sum rate maximization problem can be solved for
any systems with convex capacity regions, i.e., MIMO and OFDM
uplink/downlink channel, most of results presented in this paper
can also be applied to these systems.

\bibliographystyle{IEEEtran}
\bibliography{own_publications,queue_publications}
\end{document}